\documentclass[aps,pre,twocolumn,showpacs,showkeys]{revtex4}
\usepackage{graphicx}
\begin{document}

\title{Pressure dependence of two-level systems in disordered atomic chain}
\author{A. Shelkan}
\email{shell@fi.tartu.ee}
\author{V. Hizhnyakov}
\affiliation{Institute of Physics, University of Tartu, Riia 142,
51014 Tartu, Estonia}

\begin{abstract}

The dependence of two-level systems in disordered atomic chain on
pressure, both positive and negative was studied numerically. The
disorder was produced through the use of interatomic pair potentials
having more than one energy minimum. It was found that there exists
a correlation between the energy separation of the minima of
two-level systems $\Delta$ and the variation of this separation with
pressure. The correlation may have either positive or negative sign,
implying that the asymmetry of two-level systems may in average
increase or decrease with pressure depending on the interplay of
different interactions between atoms in disordered state. The values
of $\Delta$ depend on the sign of pressure.

\end{abstract}

\pacs{45.05.+x, 61.43.-j, 62.50.-p} 

\keywords{Disordered solids; Discrete systems; High-pressure effects} 


\maketitle

\draft

\section{Introduction}

As it was ascertained in last years a number of anomalous properties of glasses
at low temperatures can be explained supposing that there exist anomalous low-energy 
excitations: two-level (tunneling) systems (TLSs) \cite{Anderson,Phillips} 
and quasi-localized modes \cite{KarpPar,Buchenau,Krivoglaz}. A characteristic property 
of these excitations is their remarkable sensitivity to different impacts including 
external pressure $P$.

The pressure dependence of TLSs in glasses, on the basis of rather general arguments, 
was considered by Phillips \cite{Phillips1}. He concluded that the anomalous (negative) 
value of the Gr${\rm \ddot u}$neisen constant for vitreous silica may be explained 
supposing that the TLSs, when pressure is applied, have a slight tendency to become  
less asymmetric. Such a tendency means that the energy difference between the two 
potential energy minima of a TLS   $\Delta$ and the derivative of  $\Delta$ with respect 
to $P$ are to some extent correlated, so that the mean value 
$\langle \Delta \, \partial \Delta/\partial P \rangle$ differs from zero (and is negative).

In \cite{HizhKik}, it was shown that the temperature cycling effect
on the width of spectral holes, burned in a dye-doped polymer glass
at high pressure and low temperature can be understood if one
assumes that under external pressure the higher minimum in the
potential energy of TLSs gains more energy than the lower minimum.
This means that in polymer glasses there also exists a correlation
between $\Delta$ and $\partial \Delta/\partial P$; however, it is of
the opposite (i.e. positive) sign, so that TLSs in these glassy
materials become more asymmetric with increasing pressure.

Although the existence of correlation between  $ \Delta $ and
$ \partial \Delta/\partial P$  in amorphous solids is expected, it is of interest
to verify this property of disordered state by direct molecular dynamics simulation.
For checking this correlation numerically, one must bear in mind that
the concentration of TLSs in amorphous state is usually rather low and therefore
quite large clusters of atoms should be involved in computations, making them fairly
intricate and time consuming. A relatively simple opportunity here is to compute
one-dimensional (1D) disordered system consisting of a reasonably moderate
number of atoms.

In this study, we have carried out the computations of a disordered atomic chain.
To produce a disordered state, special pair potentials having more than one
energy minimum were employed. In this model, already in a single-component case,
one gets the states of the chain with interatomic bonds of different length.
In our computations we employed the pair potentials of two types:
the potential Z1 in Ref. \cite{Doye} with Friedel oscillations
and the Schilling piecewise parabolic double-well potential \cite{Schilling},
which has a discontinuity of its first derivative at the barrier separating the minima.

\section{Disordered chain with smooth multi-well pair-potential}
Let us first discuss a monatomic chain with the pair potential Z1 \cite{Doye}
\begin{equation}
V(r) = a \, e^{\alpha r}\cos{(2 k r)}/r^3+ b \left(
\sigma/r\right)^n+ V_0, \label{Vr} \label{VD}
\end{equation}
where $r$ is the distance between atoms in proper units. Values of the parameters
$a = 1.58$, $\alpha = -0.22$, $b = 4.2 \cdot 10^8$, $\sigma = 0.331$, $n =18$
are the same as those given in \cite{Doye} for the potential Z1. The quantity
$V_0$ is a constant (we take $V_0 = 0$). This potential has the sequence of minima and maxima describing the Friedel oscillations. This-type potentials are typical for metals where the oscillations describe the effect of screening of the electric field by free electrons. The parameter $k$ determining the period of oscillations is given by the Fermi wave vector. This parameter will be varied below to show that one can observe distinct kinds of response of the chain to the applied pressure.

In our computations we examined the chain consisting of 400 atoms.
We took into account the interactions between $24$ nearest atoms (inclusion of more
interactions did not noticeably change the results). The disordered state was
generated by randomly selecting 20 percent of the nearest neighbor pairs of atoms
and placing them at the distance, corresponding to the second minimum
(situated at $r \approx 1.8$) of the potential given by Eq. (1). All the other
nearest neighbor pairs were initially placed at the distance ($r \approx 1.1$),
corresponding to the first (main) minimum. The initial configuration of atoms
was allowed to relax to the local potential minimum of the chain. In order to find
the relaxed configuration, the equations of motion of atoms were integrated using
the forth-order Runge-Kutta algorithm. The motion towards the relaxed configuration
was accomplished step by step, zeroing of the velocities of all atoms at every
five to ten time steps. This procedure was repeated until only negligibly small
changes (less than  $10^{-8}$ in our units) in the positions of all atoms
were observed, indicating that the final configuration of the chain was reached.

TLSs in the present model are located at the boundaries of "dense" and "rarefied"
islands (groups of atoms with small ($r \approx 1.1$) and large ($r \approx 1.8$)
distance, respectively). The tunneling transition in a TLS corresponds to the  motion of the atom(s)
located at the boundary of short and long bonds, resulting in the permutation of two
bonds, i.e. in the displacement of a long bond to another position; see, e.g.
the changes in the fragment of the atomic chain in the picture below - displacements
of atoms (and bonds) cause the transitions between the configurations presented
in the different lines of the picture.

\vspace{0.1cm}

\begin{picture}(220,70)
\put(50,55){\circle{6}}
\put(70,55){\circle{6}}
\put(90,55){\circle{6}}
\put(120,55){\circle{6}}
\put(150,55){\circle{6}}
\put(170,55){\circle{6}}
\put(50,35){\circle{6}}
\put(70,35){\circle{6}}
\put(100,35){\circle{6}}
\put(120,35){\circle{6}}
\put(150,35){\circle{6}}
\put(170,35){\circle{6}}
\put(50,15){\circle{6}}
\put(80,15){\circle{6}}
\put(100,15){\circle{6}}
\put(120,15){\circle{6}}
\put(150,15){\circle{6}}
\put(170,15){\circle{6}}
\end{picture}

\subsection{Free ends}
Our first computation was done for the chain with free ends. In the first run,
the positions of atoms of the relaxed initial configuration were determined and
the potential energy of the chain and its length $L_0$ were found. After the first
run, one of the long bonds was displaced to one of the nearest positions
(this displacement corresponds to the single-particle tunneling transition).
Then the new coordinates of atoms in the relaxed configuration and the new length
were computed and the new potential energy of the chain found. The difference
of this energy value from the value corresponding to the initial relaxed
configuration gave us the magnitude of $\Delta$. The described runs were repeated
with displacements of all other long bonds to the nearest positions, and the
values of $\Delta$ of all other single-particle TLSs were found.

Our computations revealed that the largest $|\Delta|$   belongs to the TLSs
which correspond to the configurations  presented in the first and second lines of
the picture.  If we mark the short bond by  $1$  and the long bond by $2$,
then the back and forth tunneling transitions in these TLSs can be denoted as
(1 1 2 2 1) $\leftrightarrow$   (1 2 1 2 1).  The next largest $|\Delta|$ belongs
to the TLSs  which correspond to the configurations presented in the  second and third
lines of the picture (the tunneling transition in these TLSs can be denoted as
(1 2 1 2 1) $\leftrightarrow$ (2 1 1 2 1)). The configurations \ with \ more \ than \ two
\ adjacent \ long \ bonds, \ e.g.,\  (1 2 2 2 1), (1 2 2 2 2 1), etc., lead also to the 
TLSs with a comparable $|\Delta|$. The concentration of  these structures in our case is
the smaller the larger is the "rarefied" island. In the model under consideration,
the structures with extra long bonds corresponding to the third, the fourth, etc.
minimum of the potential given by  Eq. (\ref{VD}) can also exist. However, these
structures were ignored relying on the physical argument of their weak stability.

The energy difference $\Delta$ of a TLS depends also on the degree of disorder. This
dependence is caused by the interaction of the described above central atoms of the
TLS with the surrounding atoms. This conclusion is supported by consideration of the
chains with lower and higher concentration of long bonds. Namely, it was found that
in the first case the dispersion of the values of   $\Delta$ of the TLSs with
the same configuration of  the central atoms is diminished and in the second case
enlarged.

We have also studied the cooperative TLSs that correspond to a simultaneous
displacement of pairs and larger groups of atoms (e.g. tunneling transitions
(1 1 2 2 1 1)  $\leftrightarrow$ (2 1 1 2 1 1), (1 2 1 2 1 2) $\leftrightarrow$
(2 1 2 1 2 1), etc). It was found that the main conclusions which were drawn  above
for single-particle TLSs, hold also for cooperative TLSs. However, as more particles
have to move, the effective mass of the tunneling group of atoms is increased, which
implies rather long relaxation times. Therefore, the contribution of such TLSs
to low-temperature characteristics can be neglected on a time scale much shorter
than the corresponding relaxation time \cite{Reichert}.

\subsection{Isobaric and isochoric tunneling transitions}
The difference of the lengths of different configurations of the chain
with free ends is very small as compared to $L_0$.
The larger the $L_0$ is the smaller is the pressure and the work required
for bringing the lengths of these configurations to the same length $L_0$.
Thus, for sufficiently large $L_0$, the energies of the configurations of the chain
with free ends are practically equal to the energies of the corresponding configurations
of the chain with fixed length $L_0$.
The same is true for the corresponding values of $\Delta$.

An analogous situation exists for isobaric and isochoric tunneling transitions for nonzero
pressure. An external pressure applied to the ends
of the initial configuration of the chain causes the change of the length of the chain
from $L_0$ to $L$. In the isobaric case (the pressure $P$ is fixed) after
a tunneling transition the latter length in its turn changes from $L$ to $L+dL$.
The difference $\Delta$ of the energy of corresponding configurations includes the term
$PdL$. The work required to bring these configurations to the same length $L$,
i.e. to the isochoric case, is also $PdL$ if we assume
the change of $P$ required for bringing the lengths of these configurations
to the same length $L$ to be negligibly small.
The larger the chain is the better this assumption is fulfilled,
and the smaller is the difference between the isobaric and isochoric $\Delta$ values.
In our case of the chain of 400 atoms the difference is less than 0.1 $\%$.

Here, following \cite{Phillips1} we consider the isochoric tunneling transitions.
We take pressure $P$ in reduced dimensionless units according
to the relation $P \equiv (L_0-L)/ L_0$. The case $P=0$ corresponds to the free ends of
the initial (i.e. prior to the tunneling transition) configuration of the chain.

\subsection{Effect of compression}
In the next step the effect of positive pressure (in our case a uniaxial
compression of the chain) was examined. Here the positions of
the end atoms were fixed so that the length of the
chain  $L$ would be the same (in the $P=0$ case) or shorter (in $P>0$ case)
than the length $L_0$ of the initial relaxed
configuration of the chain with free ends. The initial positions of all other atoms
were chosen so that the distances between atoms would be uniformly reduced.
By using this initial condition, the relaxed positions of
all by end atoms were computed and the potential energy of the chain found.
Then in a similar manner as in the runs without pressure, one of the long bonds was
displaced to another (nearest) position, the new relaxed configuration was computed
and the new potential energy found. The difference between this energy value and the
value corresponding to the initial relaxed configuration gave us the energy
difference of the two minima of a TLS at a pressure  $P$, denoted by $\Delta (P)$.
The described runs were also repeated with displacements of all other long bonds,
and the values of $\Delta (P)$ of all other TLSs were found.

Figures 1 and 2 show the obtained dependence of $\Delta (P) - \Delta (0)$
on $\Delta(0)$ for different TLSs at a small pressure $P = 0.001$ in the case
of the potential Z1 given by Eq. (\ref{VD}) with the parameters $k=4.12$ and $k=4.5$,
respectively.
From these figures it is clearly seen that  $\Delta (0)$ and
$\Delta (P) - \Delta (0)$ are correlated. The sign of the correlation  may be
different: in the case  $k=4.12$ (Fig. 1) the correlation is positive (TLSs become more
asymmetric under pressure), in the case $k=4.5$ (Fig. 2) it is negative. The correlation
is almost complete.  However, there exist intermediate cases (with overall weak
dependence of $\Delta$ on $P$) where the correlation is insignificant (see, e.g.,
Fig. 3).
\begin{figure}[th]
\begin{center}
\includegraphics*[angle=-90,width=.45\textwidth]{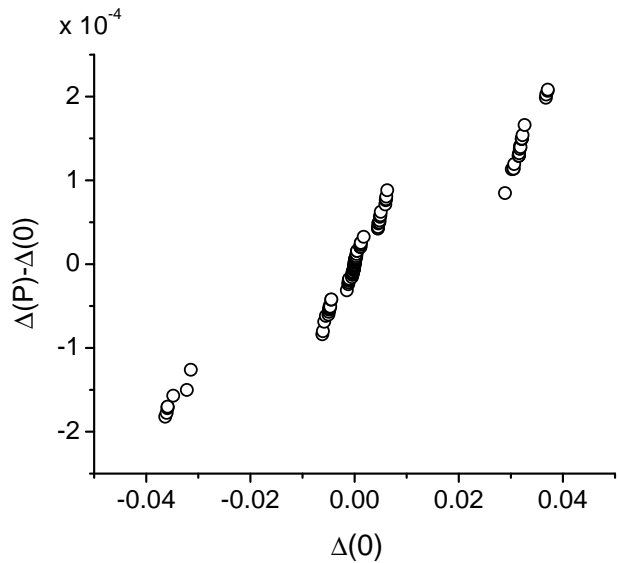}
\end{center}
\caption{Changes in  $\Delta$ of TLSs under dimensionless pressure $P =0.001$
in the disordered atomic chain with the pair potential Z1. The presented case
corresponds to k = 4.12 in Eq. (\ref{VD}); the other parameters are given in the text.} 
\end{figure}

\begin{figure}[th]
\begin{center}
\includegraphics[angle=-90,width=.45\textwidth]{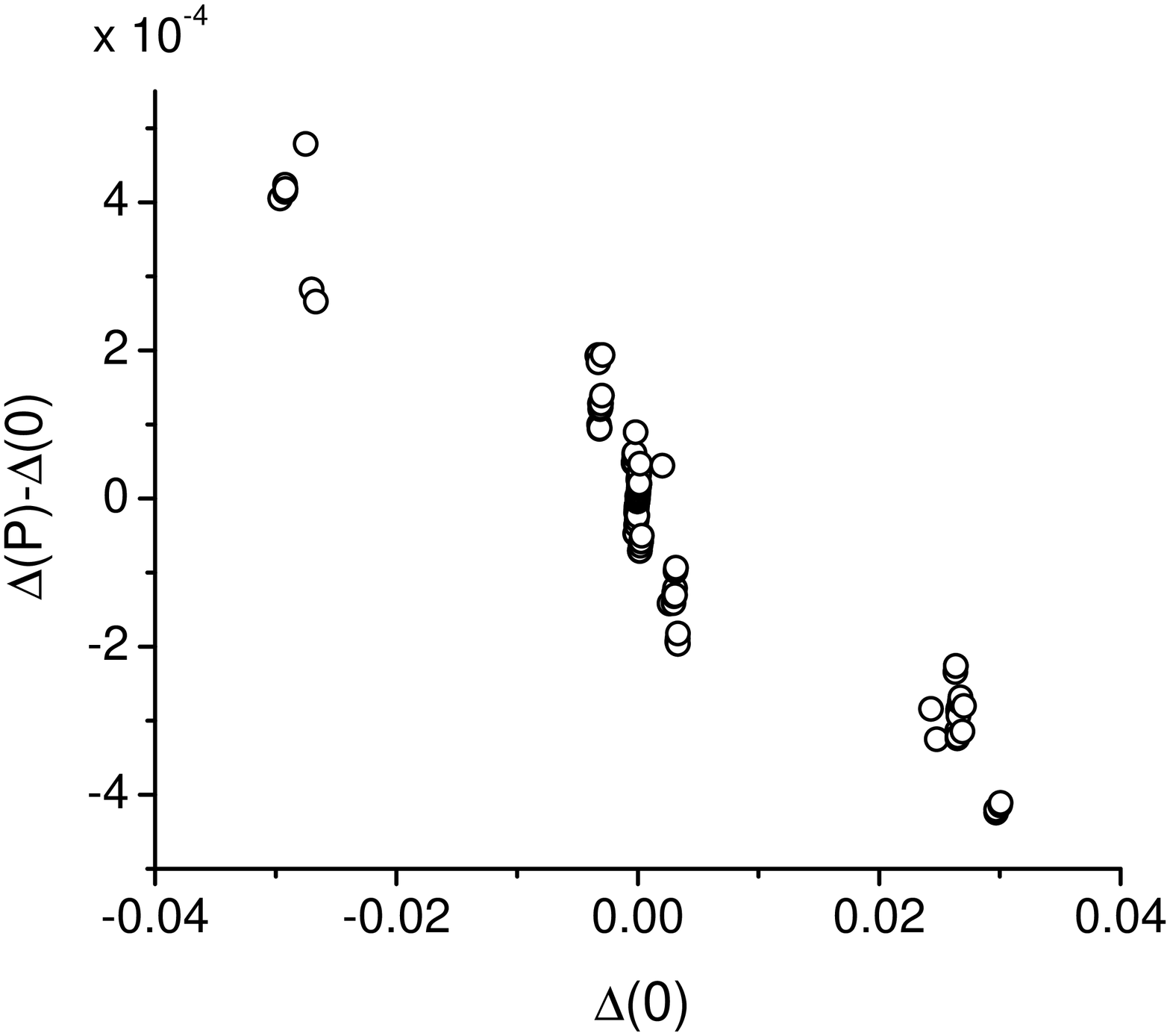}\hspace*{0em}
\end{center}
\caption{The same as in Fig. 1, but for $k=4.5$.}
\end{figure}

\begin{figure}[th]
\begin{center}
\includegraphics[angle=-90,width=.42\textwidth]{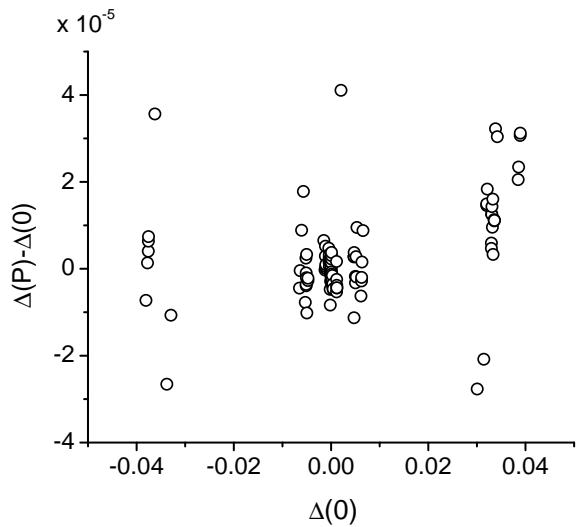}\hspace*{0em}
\end{center}
\caption{The same as in Fig. 1, but for $k=4.3$.}
\end{figure}

Note further that in the case $k=4.12$ the minimum of a TLS with a higher energy
corresponds to a larger length of the chain with free ends. However in the case
$k=4.5$ the relation between energy and length is just the opposite: the minimum
with larger energy corresponds to a smaller length. In both cases the larger is the
length, the bigger is the change (increasing) of the energy with pressure.
This can be understood as follows - to compress the chain extra energy is needed,
which is the larger the larger is compression.

Unlike the case of low $P$ values, in the case of high $P$ values the correlation
between $\Delta$  and \ $\partial \Delta / \partial P$ \  was found to be always
positive. In the region of high enough pressures the TLSs are more asymmetric than
at zero pressure. It is clearly seen in Fig. 4, where the dependence of the energy
differences \, $\Delta$ \, of two \, TLSs \, of the types  (1 1 2 2 1) $\leftrightarrow$
(1 2 1 2 1) and (1 2 1 2 1) $\leftrightarrow$ (2 1 1 2 1) on dimensionless pressure
$P$ is given for the chain with the same interaction parameters as in Figs. 1 and 2.
The upper values of $P$ in Fig. 4 correspond to the critical pressure at which the
abrupt changes in a chain begin to occur. Such changes take place when some long bonds
transform into the short bonds, which results in disappearance of respective TLSs.
Obviously, this happens in such a manner that TLSs gradually increase their asymmetry.
With further increase of pressure more and more TLSs disappear. At  sufficiently high
pressure $P \sim 0.1$ there remain no TLSs at all and the chain becomes ordered.

The presented results are in agreement with the experiments
\cite{Anders,PRB,Eller}, where a reduction in the number of TLSs and
soft localized modes with pressure in glasses was observed.
The reduction in the number of states of a disordered
structure with the growth of its density was earlier found
theoretically in \cite{Hanner} by an analytical study of a chain of
particles with Schilling piecewise parabolic double-well potential
\cite{Schilling}. An analogous result was obtained in \cite{Weber}
by the molecular dynamics simulation of 32 and 108 particles with
the finite-range Lennard-Jones interaction between them.

\begin{figure}[th]
\begin{center}
\includegraphics[angle=-90,width=.48\textwidth]{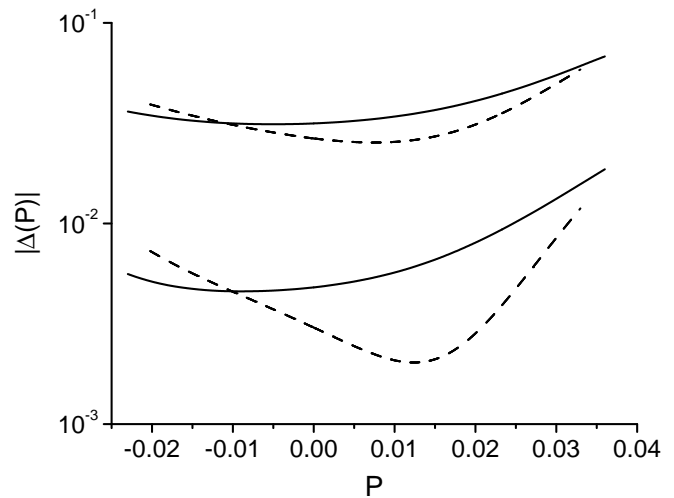}\hspace*{0em}
\end{center}
\caption{Pressure dependence of $|\Delta|$ in the disordered atomic chain for the
two TLSs with the largest $|\Delta|$ values. Pair potentials are the same
as in Figs. 1 and 2; solid lines correspond to $k=4.12$ while dashed lines to $k=4.5$.}
\end{figure}

\subsection{Effect of stretching}
It is reasonable to expect that the response
of a glass to the volume contraction and dilatation should be
different \cite{Taka,HizhEst}. If so then the value of $\Delta(P) $
should depend on the sign of $P$. In order to control this statement
we performed computations of TLSs also for negative $P$ (in our case
for uniaxial stretching of the chain). We have found that, indeed, the value of
the energy $\Delta$ of a TLS depends on the sign of pressure, which
can clearly be seen in \ Fig. 4. Only \ in \ the case \ of \ a large
\ $|P|$ \ the change of \ $\Delta$ \ with \ $|P|$ \, for \ positive \ and \ negative
\ $P$ \ is \ similar: in both cases $\Delta$ increases with $|P|$. It was also found 
that at negative pressure $P \approx -0.02$ the chain breaks down.

\section{Disordered chain with piecewise parabolic double-well potential}
Lastly we have performed computations of the pressure dependence of 1D disordered chain
with the Schilling-type piecewise parabolic double-well potential for nearest
neighbors \cite{Schilling,Hanner}
\begin{equation}
V_1(r)\!=V_0 + \!a_1(r-r_1)^2 \Theta(r_0-r)+[a_2(r-r_2)^2+ b] \Theta(r-r_0),\! \label{VSh}
\end{equation}
where $\Theta (x)$ is the Heaviside step function. In \cite{Schilling,Hanner} only
the case $a_1=a_2$ was considered. Here the values of parameters $V_0=-0.743$, $a_1 = 40$,
$a_2 = 5.4$, $b = 0.6$, $r_0 = 1.357$, $r_1=1.13$, $r_2=1.88$ were chosen such that
the minima of the potential given by Eq. (\ref{VSh}) would coincide with the first
two minima of the used above potential Z1. In the model \cite{Schilling,Hanner},
besides the interaction between nearest neighbors, also the interaction with
next-nearest neighbors is taken into account. This interaction is represented by
a parabolic potential
\begin{equation}
V_2(r)=c(r-r_3)^2.
\label{VSh2}
\end{equation}
We took $c = 0.05$ or $c=-0.05$ and $r_3= 4.0$. The results of computations of TLSs
for small dimensionless pressure $P=0.001$ are shown in Fig. 5. It is clearly seen
that the quantities  $\Delta(0)$ and $\Delta (P)- \Delta (0)$ are correlated, and
the correlation is complete and negative. The variance of the TLSs parameters
is very small - much smaller than in the case of the potential Z1. This
is a consequence of a strong localization of the Schilling interactions.

\begin{figure}[th]
\begin{center}
\includegraphics[angle=-90,width=.44\textwidth]{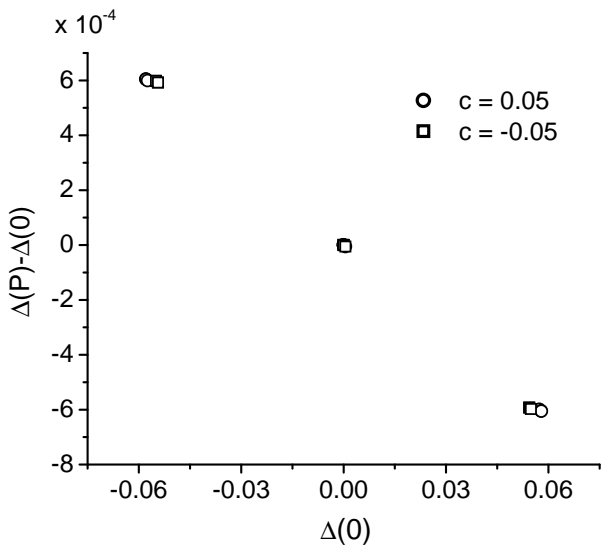}\hspace*{0em}
\end{center}
\caption{Changes in $\Delta$ under dimensionless pressure $P=0.001$ in
a disordered chain with the Schilling-type pair potential. The presented cases correspond
to $c=0.05$ and $c=-0.05$ in Eq.(\ref{VSh2}); the other parameters are given in the
text.}
\end{figure}

\vspace{2cm}

We have also performed computations for other values of the parameters of the
potentials $V_1(r)$ and $V_2(r)$. It appeared that the sign of the difference
$a_2 - a_1$ determines the sign of the correlation between $\Delta$ and
$\partial \Delta /\partial P$: if $a_1 > a_2$ then the sign of the correlation is
negative, thereby if $a_1 < a_2$ then the sign is positive. In the case $a_1=a_2$
the values of $\Delta$ do not depend on $P$. In the large $P$ limit all the TLSs
disappear.

\section{Conclusion}
Molecular dynamics simulations of 1D disordered state unambiguously
testify that for two-level systems there exists a correlation between the magnitudes
of $\Delta$  and their alterations under external pressure. The correlation may be
either positive or negative sign. This signifies that the asymmetry of two-level
systems may in average increase or decrease with pressure, depending on the interplay
of interactions between particles in the disordered state.
The response of TLSs to external pressure depends on the sign of pressure.

\section{Acknowledgement}


The research was supported by the Estonian Science Foundation through Grant No. 7741.
The authors are grateful to Prof. J. Kikas and Dr. A. Laisaar for valuable
discussions.


\end{document}